\documentclass[useAMS,usenatbib]
{mn2e}
\usepackage{epsfig}
\usepackage{amsmath, amssymb,bm}

\def \be{\begin{equation}}
\def \ee{\end{equation}}

\def \msun{\rm M_{\odot}}

\def  \le{{L_{\rm Edd}}}
 
\begin{document}
\title[]{ The Perils of Pdot}

\author[Andrew King \& Jean--Pierre Lasota] 
{\parbox{5in}{Andrew King$^{1, 2, 3}$ \& Jean--Pierre Lasota$^{4, 5}$
}
\vspace{0.1in} \\ $^1$ School of Physics \& Astronomy, University
of Leicester, Leicester LE1 7RH UK\\ 
$^2$ Astronomical Institute Anton Pannekoek, University of Amsterdam, Science Park 904, NL-1098 XH Amsterdam, The Netherlands \\
$^{3}$ Leiden Observatory, Leiden University, Niels Bohrweg 2, NL-2333 CA Leiden, The Netherlands\\
$^{4}$ Institut d'Astrophysique de Paris, CNRS et Sorbonne Universit\'e, UMR 7095, 98bis Bd Arago, 75014 Paris, France\\  
$^{5}$ Nicolaus Copernicus Astronomical Center, Polish Academy of Sciences, ul. Bartycka 18, 00-716 Warsaw, Poland\\     
}


\maketitle

\begin{abstract}
\citet{S24} has recently published observations of binary period derivatives $\dot P$ for 52 cataclysmic variables, and concluded that these strongly conflict with all proposed evolutionary pictures for these systems. We point out once again that using measurements of $\dot P$ is likely in practice to produce misleading evolutionary constraints in almost every case. The one identified exception to this is probably the recently--born X--ray binary SN 2022jli, because of its extremely high mass transfer rate.
\end{abstract}


\footnotetext[1]{E-mail: ark@astro.le.ac.uk}
\citet{S24} has recently published observations of the rate $\dot P$ of binary period change $\dot P$ in 52 cataclysmic variable systems (CVs). The measured values of 
$\dot P$ do not conform to those expected in the standard picture of CV orbital evolution under long--term angular momentum loss via gravitational radiation and magnetic braking. This relates the long--term value of $\dot P$ to that of the mass transfer rate $-\dot M_2$ from the donor star (see e.g. King, 1988 for a review). 

\citet{S24} concludes that the reported observations invalidate this evolutionary picture. But a very long line of papers --  most recently \citet{KL21} in another context -- have reiterated the fundamental point that stars do not have sharp edges, so cannot follow the predicted evolutionary relations on arbitrarily short timescales. The predicted relations between $\dot P$ and  $-\dot M_2$ hold  only when averaged over timescales significantly longer than the timescale $t_H$ on which the angular momentum losses move the Roche lobe one density scaleheight $H$ through the donor star. In almost every case, this requires observations over unfeasibly long timescales. 

Standard formulae show that the scaleheight for a CV donor star of radius $R_2$ is given by
\begin{equation}
\frac{H}{R_2} \sim \frac{R_2 kT_{\rm eff}}{GM_2\mu m_H} \sim 3\times 10^{-4},
\end{equation}
where the star's effective temperature $T_{\rm eff}\sim 3000 - 5000\ {\rm K}$. Then 
\begin{equation}
t_H \sim \frac{H}{\dot R_2} \sim \frac{H}{R_2}\frac{M_2}
{|\dot M_2|} \sim 3\times 10^5\,{\rm yr}
\label{tH}
\end{equation}
for a CV evolving on a timescale $\sim M_2/|\dot M_2|\sim 10^9$~yr (e.g. under magnetic braking), or $\sim 10$ times longer for evolution under gravitational radiation. 
In a monitoring programme lasting 
$\sim 30$~yr, the Roche lobe will typically have moved by 2 metres, compared with the density scaleheight $\sim 200$~km.

Since only observations on timescales $> t_H$ can constrain the long--term evolution of CVs, we conclude that those reported by \citet{S24} do not pose problems for the currently--accepted evolutionary picture. The observed period changes must be dominated by short--term phenomena, so it is unsurprising that they frequently show alternating or unexpected signs for example.

The form of Eq. (\ref{tH}) shows that the tempting idea that one might measure a mean mass transfer rate in a binary system by the much easier method of measuring 
$\dot P$ is unlikely to work very often in practice. Evidently this requires a much higher mass transfer rate $-\dot M_2$ than in CVs. To our knowledge the only case where this appears to occur is in the newly--born X--ray binary system SN 2022jli (\citealt{Chen24,Moore23}). Here the very recent supernova explosion producing the compact accretor has violently disturbed the binary geometry and forced the companion to overflow its Roche lobe very significantly, producing a mass transfer rate $\gtrsim 10^{-5}\msun\, {\rm yr}^{-1}$ (King \& Lasota 2024). This gives $t_H \sim 1\, {\rm yr}$, and probably accounts for the observed luminosity decay on a timescale 
$\sim 250$ days.


\section*{Acknowledgments}

We thank Jim Pringle for discussions.

{}

\end{document}